\documentclass{pasj00} 
\begin{document}
\SetRunningHead{Y. Sofue, et al.}{Unified Rotation Curve of the Milky Way Galaxy}
\Received{2008/mm/dd}  \Accepted{2008/mm/dd} 

\def\kms{km s$^{-1}$}  
\def\Msun{M_\odot}
\def\be{\begin{equation}}
\def\ee{\end{equation}}
\def\bc{\begin{center}}
\def\ec{\end{center}}
\def\dv{de Vaucouleurs}

\title{Unified Rotation Curve of the Galaxy --- Decomposition into de Vaucouleurs Bulge, Disk, Dark Halo, and the 9-kpc Rotation Dip ---}
\author{Yoshiaki {\sc Sofue}$^{1,2}$, Mareki {\sc Honma}$^3$, and Toshihiro {\sc Omodaka}$^1$ }  
\affil{
1. Department of Physics and Astronomy, Kagoshima University, Kagoshima, 890-0065 Kagoshima\\
2. Institute of Astronomy, University of Tokyo, Mitaka, 181-0015 Tokyo, \\
3. Mizusawa VERA observatory, National Astronomical observatory of Japan,\\
 Graduate University for Advanced Study, Mitaka, Tokyo 181-8588  \\
Email:{\it sofue@sci.kagoshima-u.ac.jp and sofue@ioa.s.u-tokyo.ac.jp}
 }


\maketitle

\begin{abstract} 
We present a unified rotation curve of the Galaxy re-constructed from the existing data by re-calculating the distances and velocities for a set of galactic constants $R_0=8$ kpc and $V_0=200$ \kms. We decompose it into a bulge with \dv-law profile of half-mass scale radius 0.5 kpc and mass $1.8\times 10^{10}\Msun$, an exponential disk of scale radius 3.5 kpc of $6.5 \times 10^{10}\Msun$, and an isothermal dark halo of terminal velocity 200 \kms. The $r^{1/4}$-law fit was obtained for the first time for the Milky Way's rotation curve. After fitting by these fundamental structures, two local minima, or the dips, of rotation velocity are prominent at radii 3 and 9 kpc. The 3-kpc dip is consistent with the observed bar. It is alternatively explained by a massive ring with the density maximum at radius 4 kpc. The 9-kpc dip is clearly exhibited as the most peculiar feature in the galactic rotation curve. We explain it by a massive ring of amplitude as large as 0.3 to 0.4 times the disk density with the density peak at radius 11 kpc. This great ring may be related to the Perseus arm, while no peculiar feature of HI-gas is associated.
\end{abstract}

\section{Introduction}

Rotation curve is the fundamental tool to derive the mass distribution in the Galaxy. A great deal of data for the galactic rotation curve have been obtained by various methods, such as using terminal velocities of the HI and CO lines for the inner Galaxy (Burton, Gordon 1978; Clemens 1985; Fich et al. 1989); optical distances and CO-line and optical velocities for outer disk (Blitz et al. 1986; Demers and Battinelli 2007); HI thickness for the entire disk (Merrifield 1992; Honma and Sofue 1997a,b); and recently parallax and proper motion measurements using VERA for the outer disk (Honma et al. 2007). 

However, the entire rotation curve is still crude because of the large scatter in the data, not only because of the different methods, but also for the different parameters used to convert the observed data to rotation velocities.  This has made it difficult to compare among the existing rotating curves as well as to compare them with model calculations. In this paper, we unify the existing data into a single rotation curve by re-calculating the distances and velocities adopting a nominal set of the galactocentric distance and the circular velocity of the Sun as  $(R_0, V_0)$=(8.0 kpc, 200 \kms). 

The mass distribution has been obtained by decomposition of a rotation curve into several components such as a bulge, disk and dark halo (e.g., Bosma 1981;  Kent 1986; Sofue 1996). It is well established that the luminosity profile of the spheroidal component in galaxies is represented by the \dv\ ($e^{-r^{1/4}}$ or $r^{1/4}$law; 1953, 1958) law with some modifications to $r^n$ law including asymmetries (Ciotti et al. 1991; Trujillo et al. 2001).  Decomposition using the \dv\  and S{\'e}rsic ($r^n$) profiles have been applied to early type galaxies (Noordermeer 2008). However, there has been no attempt to fit the Milky Way's rotation curve by the \dv\ law. The \dv\ law is known for its steep rise of density toward the center, and indicating that the volume density increases to infinity at the nucleus. This is particularly important for the relationship of the bulge to the central massive black hole (Kormendy and Richstone 1995; Melia and Falcke 2001). 

After fitting the newly developed rotation curve by the \dv\ bulge, exponential disk, and an isothermal dark halo, we discuss the two prominent rotation dips at radii 3 and 9 kpc. Particularly, the 9-kpc dip is exhibited as the most pronounced and peculiar feature in the Galactic rotation curve.  We discuss the dips in relation to a bar, wavy rings, and/or spiral arms. 

\section{Unified Rotation Curve of the Galaxy}
 
We create an updated rotation curve for the galactic constants $R_0=8.0$ kpc and $V_0=200$ \kms by integrating the existing data from the literature, and plot them in the same scale. The adopted galactocentric distance is most widely used in the literature (e.g. Ghez et al. 2005), but is smaller than the IAU-recommended value of 8.5 kpc, and larger than the most recent value of about 7.5 kpc (e.g. Nishiyama et al. 2006; Groenewegen et al. 2008).

The tangent point data for HI line (Burton and Gordon 1978) in the Galactic Center region and the CO-line tangent point data inside the Solar circle, both compiled and plotted by Clemens (1985), are corrected for by
\be
R_c=R_i+\Delta R_0 {R \over R_0} \label{radcor}
\ee
and
\be
V_c(R)=V_i +\Delta V(R),
\ee
where
\be
\Delta V(R)= \Delta V_0 {R \over R_0}(1+O(\Delta R_0/R_0)) 
\simeq \Delta V_0 {R \over R_0}.
\label{Vcor}
\ee
Here, $\Delta R_0=-0.5$ kpc is the correction for the solar distance, $\Delta V_0=-20$ \kms is the correction for the solar velocity, to those used in Clemens (1985), $R_i$ and $V_i$ are the radii and rotation velocities read from the literature, and $R_c$ and $V_c$ are the corrected values. 

The HI tangent point data for $l=15^\circ$ to 90$^\circ$ and $l=270^\circ$ to $345^\circ$ presented by Fich et al. (1989) were used to re-calculate the radii and rotation velocities. We also re-calculated rotation velocities and galactocentric distances of HII regions with known distances using the catalogue compiled in Fich et al. (1989), where we excluded those at galactic longitudes between $170^\circ$ and $190^\circ$ in order to avoid larger errors occurring from $1/{\rm sin}~l$ effect. Rotation velocities from the HI-disk thickness method, obtained by Honma and Sofue (1997a,b), are also plotted for the case of $V_\odot=200$ \kms. The recent optical measurements of C stars from Demers and Battinelli (2007) are also plotted, where we recalculated the values for $(R_0, V_0)=(8.0$ kpc, 200.0 \kms). Finally, we plot the latest, most accurate data at $R=13.1$ kpc from VERA observations (Honma et al. 2007) by the big circle.

\begin{figure*}
\begin{center} 
\includegraphics[width=16cm]{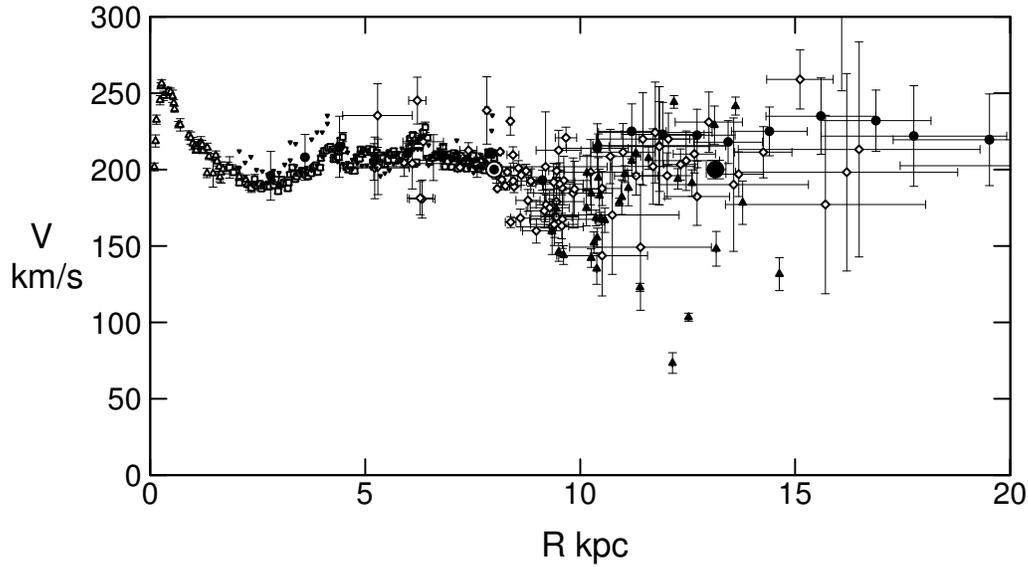}   
\end{center}
\caption{Observed circular velocities representing the rotation curve of the Galaxy. Open triangles: HI tangent velocity method (Burton and Gordon 1978); Rectangles: CO tangent (Clemens 1989); Reverse triangles: HI tangent (Fich et al. 1989); Diamonds: CO and HII regions (Fich et al.1989, Blitz et al. 1982); filled triangles: Demers and Battinelli (2007); Circles:  HI thickness (Honma and Sofue 1997a,b); Big circle at 13.1 kpc: VERA-parallax, proper motion and velocity (Honma et al. 2007).  All data have been converted to $(R_0,V_0)=(8.0$, 200.0 \kms). The plotted data are in table \ref{tab_obs}.} 
\label{fig-obs}
\end{figure*}

\begin{table}
\caption{Re-calculated data for galactic rotation curve 
in figure \ref{fig-obs}}
\begin{center}
\begin{tabular}{ll}
\hline\hline  
\\
http://www.ioa.s.u-tokyo.ac.jp/$\sim$sofue/mw/rc2009/
\\
\hline \\ 
\end{tabular}
\label{tab_obs}
\end{center}
\end{table}

Figure \ref{fig-obs} shows the obtained unified rotation curve of the Galaxy. The plotted data, shown as table \ref{tab_obs}, are available in a digitized form from our URL,
http://www.ioa.s.u-tokyo.ac.jp/$\sim$sofue/mw/rc2009/ .


The unified rotation curve shows clearly the three dominant components: the galactic bulge, disk, and outer flat rotation due to the dark halo. In addition, we recognize a prominent dip at $R\sim 9$ kpc from the flatness. This local dip can be recognized in each plot of individual observations. It has been recognized already in the rotation curve obtained from the HI thickness method (Honma and Sofue 1997a), which utilizes the entire HI disk. Individual plots for the C-stars and HII region data also indicate the same dip, as indicated by different symbols in the figure. Hence, we may consider that the dip is not due to local deviations of nearby stars from the circular motion, but it is a large scale phenomenon existing in azimuthally averaged data over large spatial coverage.  

Although the 9 kpc dip is found both in C and OB stars plots, individually, we may note that the C stars appear to trace lower velocities. Also, C stars show larger scatter. This fact might indicate that the C-stars have a higher velocity dispersion, and hence a lower circular velocity. However, it is difficult to evaluate the contribution of the velocity dispersion from the current data, because the uncertainties in distance measurements are still large. The pressure (velocity dispersion) support of the galactic disk, e.g. of C stars, would be an interesting subject for the future observations and reductions.


\section{Model Rotation Curves}

A rotation curve of a system, which is composed of a spherical bulge, disk, and a dark  halo, is calculated by
\be
V(R)^2=V_b(R)^2+V_d(R)^2+V_h(R)^2.
\ee
We fit an observed rotation curve by the calculated rotation curve $V(R)$. We adjust the parameters involved in the expressions of the individual mass components such as the masses and scale radii in order to minimize the residual between the calculation and observation. Considering that the accumulated data are not uniform, the fitting was obtained by eye-estimates after trial and error, comparing the calculated curves with the plots. 

Although the calculated rotation curve based on the three basic components approximately represents the observation, we see a considerable discrepancy between these curves at $R\sim 3$ and 9  kpc. Observed velocities show the so called 'dips' at $R\sim 3$ and 9 kpc. Such small scale variation of rotation speed cannot be attributed to the basic mass distributions, but indicates local enhancements and/or depression of the surface mass density, such as due to arms, rings, and/or a bar. 

Recent high accuracy measurement by Honma et al. (2007) using VERA, an exact data point was given on the outer rotation curve as  $(R, V(R))=(13.15 \pm 0.22 {\rm kpc}, 200\pm 6 {\rm km~s^{-1}}$. We, therefore, fit this point with some ignorance of the other data points that had higher dispersion at $R>10$ kpc.  

Since the gas mass density is two orders of magnitude smaller than the disk and background mass density, the gas disk does not influence the rotation curve at al. However, it may happen that the gas density profile reflects some more massive underlying structures, superposed on the disk. In order to examine such a case, we added a pseudo gas disk whose surface mass density is ten times that of the observed density.  The result was, however, not satisfactory, except that some wavy pattern is obtained, while their amplitudes and phases are not coincident with the observations.

Finally, we try a case with two wavy rings, and show that the model can well reproduce the observed 3 and 9-kpc dips quite well. This behavior may be compared with a model considering a bar, where the 3 kpc dip is qualitatively reproduced, but we show that the amplitude is not reproduced. This is because that the bar is a radial perturbation of mode 2, whereas the ring is a local and radial perturbation yielding a rapider change of density and potential gradients.

\section{Galactic Mass Components}

For constructing the model rotation curves, we used fundamental galactic mass components, which are the bulge, disk, and halo. We also introduced some perturbations representing the discrepancies between the observations and calculated fundamental curves. We describe individual components below.

\subsection{Bulge}

\begin{figure} 
\bc
\includegraphics[width=7cm]{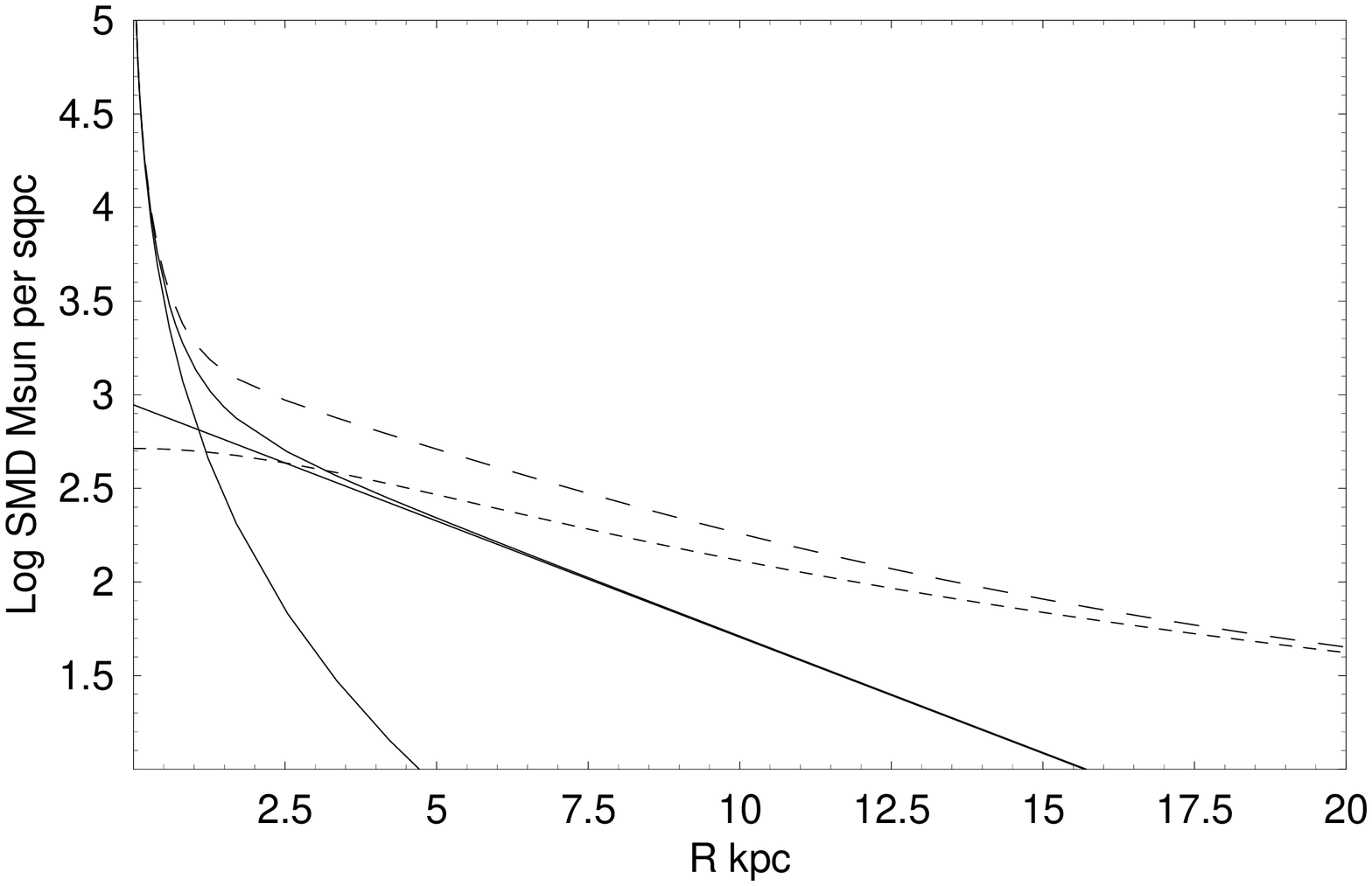}  \\
\includegraphics[width=7cm]{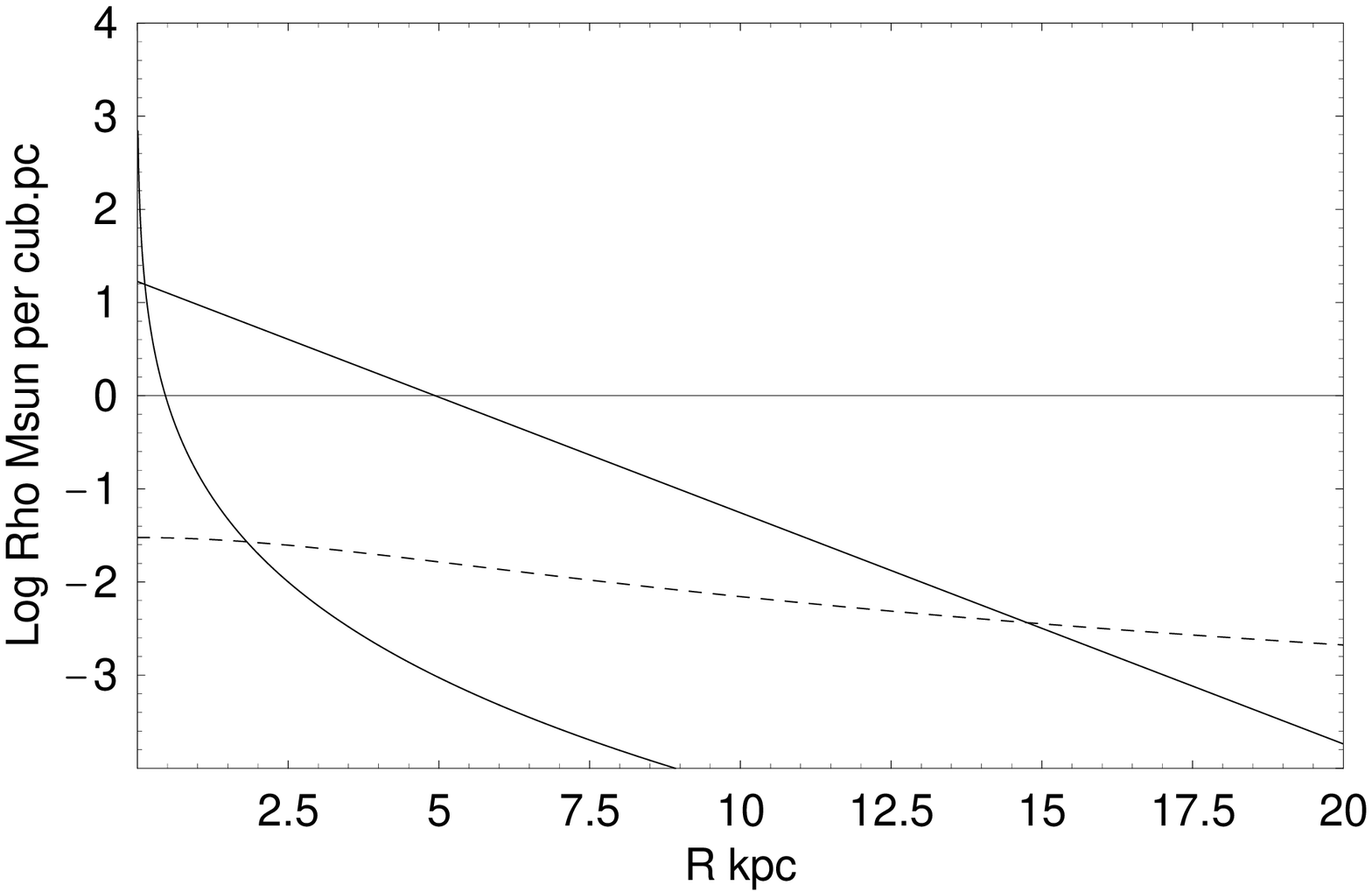}  \\
\includegraphics[width=8cm]{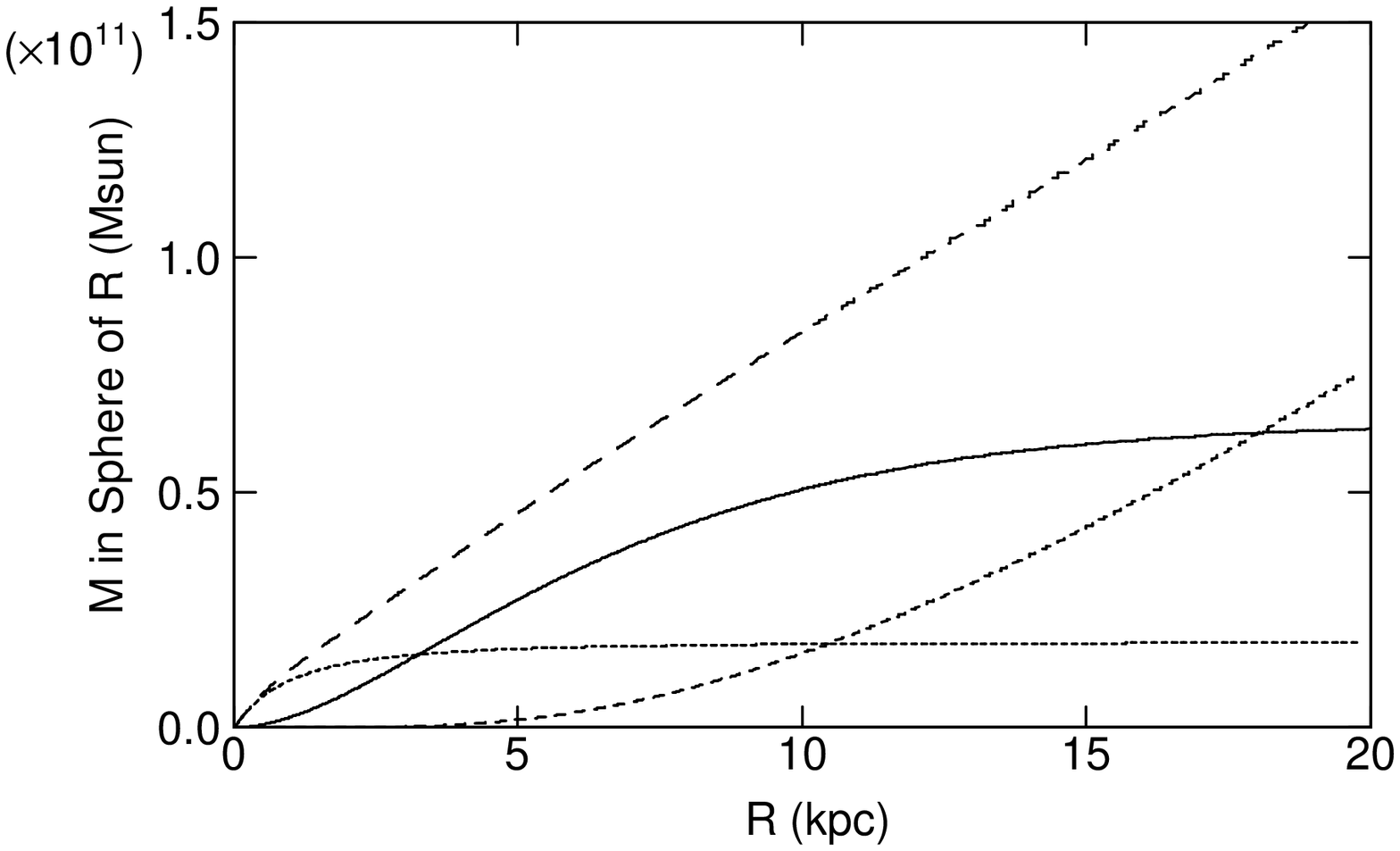} 
\ec
\caption{[Top panel]: Surface mass density distribution for the bulge, disk, dark halo, and total for a model Galaxy. High density at the center shows the bulge component with the center value of $6.8\times10^6\Msun{\rm pc}^{-2}$. Straight line indicates the exponential disk. The dashed line represents the dark halo integrated within height $-10 <z<10$ kpc. The uppermost long-dashed line is their sum. [Middle]: Volume density profile. The disk density was calculated by $\rho_d=\Sigma_d/2z$ with $z=z_0 {\rm exp}(R/R_d)/{\rm exp}(R_0/R_d)$ being the scale height and $z_0=247$ pc in the solar vicinity (Kent et al. 1991). [Lower panel]: Total masses of individual components integrated in a sphere of radius $R$. Thin line: bulge; thick solid line: disk; dash: halo;  and long dashed line: their sum.}
\label{fig-smd}  
\end{figure}

The inner region of the galaxy is assumed to be composed of two luminous components, which are a bulge and disk (Wyse et al. 1997) . The mass-to-luminosity ratio within each component is assumed to be constant, so that the mass density distribution has the same profile. The bulge is assumed to have a spherically symmetric mass distribution, whose surface mass density obeys the \dv\ law, as shown in figure \ref{fig-smd}. 

The \dv\ (1958) law for the surface brightness profile as a function of the projected radius $r$ is expressed by
\be
{\rm log} \beta = - \gamma(\alpha^{1/4}-1),
\ee
with  $\gamma=3.3308$. Here, $ \beta=B_b(r)/B_{be}$,  $ \alpha=r/R_b $, and $B_b(r)$ is the brightness distribution normalized by $B_{be}$, which is the brightness at radius $R_b$.  
We adopt the same \dv\ profile for the surface mass density:
\be \Sigma_b(r)=\lambda_b B_b(r)= \Sigma_{be} {\rm exp} \left[-\kappa \left(\left(r \over R_b \right)^{1/4}-1\right)\right]
\ee 
with $ \Sigma_{bc} = 2142.0 \Sigma_{be} $ for $\kappa=\gamma {\rm ln} 10=7.6695$.  
Here, $\lambda_b$ is the mass-to-luminosity ratio, which is assumed to be constant within a bulge.  
The total mass is calculated by
\be M_{bt}= 2 \pi \int_0^\infty r \Sigma_b(r) dr =\eta R_b^2 \Sigma_{be},
\ee
where $\eta=22.665$ is a dimensionless constant. By definition a half of the total projected mass (luminosity) is equal to that inside a cylinder of radius $R_b$.

We here adopt a spherical bulge. In fact the differences among circular velocities are not so significant for minor-to-major axis ratios greater than $\sim0.5$ (Noordermeer 2008). The volume mass density $\rho(r)$ at radius $r$ for a spherical bulge is calculated by using the surface density distribution as (Binney and Tremaine 1987; Noordermeer 2008), 
\be
\rho(r) = {1 \over \pi} \int_r^{\infty} {d \Sigma_b(x) \over dx} {1 \over \sqrt{x^2-r^2}}dx.
\ee 
Since the mass distribution   is assumed to be spherical, the total mass enclosed within a sphere of radius $R$ is calculated by using $rho(r)$ and the circular velocity as $ V_b(R) = \sqrt{ GM_b(R) / R} $.
Obviously, the velocity approaches the Keplerian-law value at radii sufficiently greater than the scale radius. The shape of the rotation curve is similar to each other for varying total mass and scale radius. For a given scale radius, the peak velocity varies proportionally to a square root of the mass. For a fixed total mass, the peak-velocity position moves inversely proportionally to the scale radius, or along a Keplerian line.  

\begin{figure} \bc
\includegraphics[width=9cm]{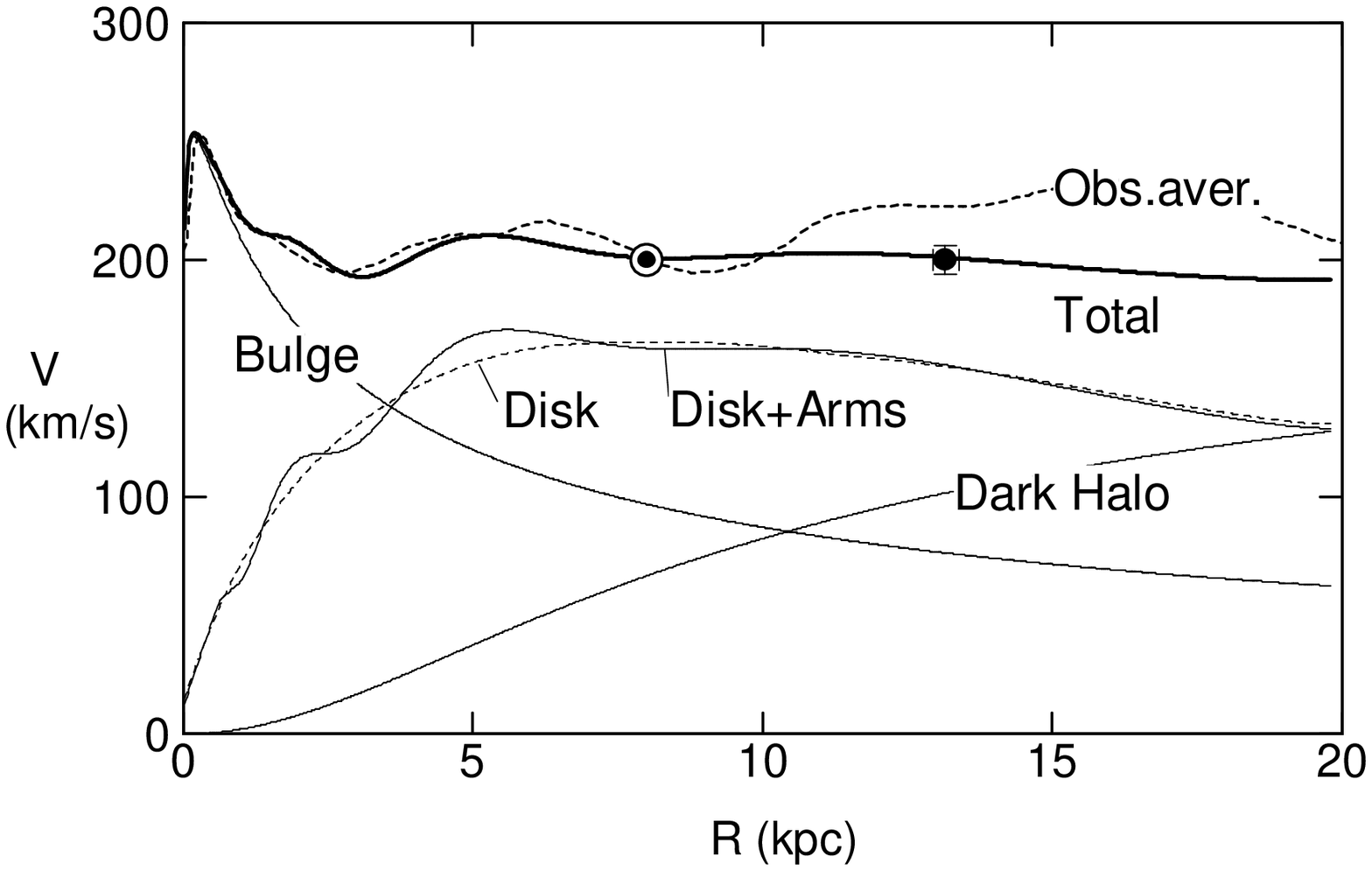}  \ec
\caption{Composite rotation curve including the bulge, disk, spiral arms, and  dark halo. The big dot denotes the observed result from VERA (Honma et al. 2007). The pure disk component is also indicated by the thin dashed line. The thick dashed line indicates a simply averaged observed rotation curve taken from Sofue et al. (1999) where the outer curve is based only on the HI data of Honma and Sofue (1997a). } 
\label{fig-rc-arm}  

\begin{center}  
\includegraphics[width=9cm]{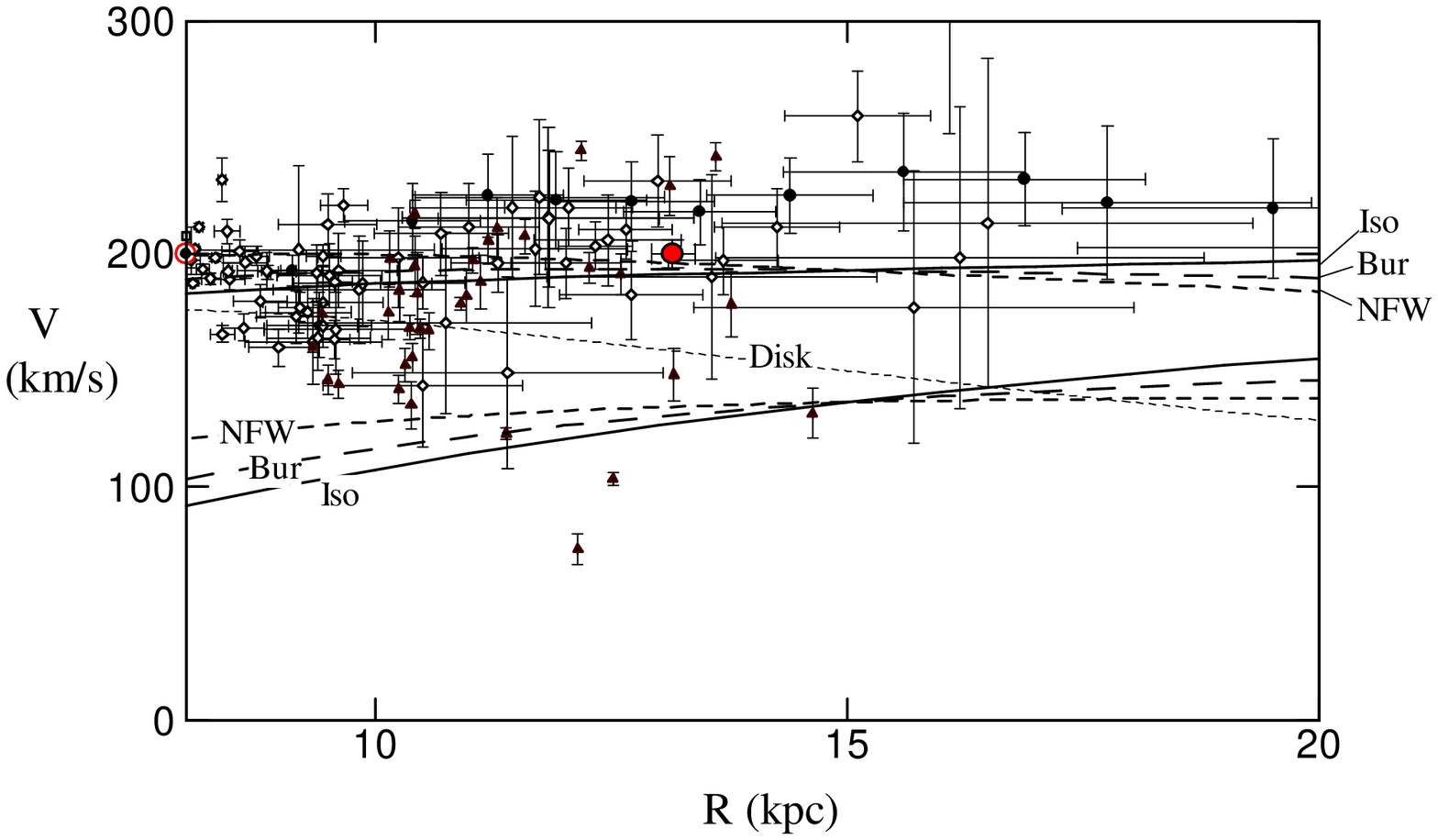}
\end{center}
\caption{Model rotation curves for three halo models: isothermal (full line), Burkert (long dash) and NFW (dashed line) models, compared with the observations. Here, we show only the disk and halo contributions, but the bulge is not added. The curves are normalized to the same value at 15 kpc. Note the large scatter in the observations and weak dependency of the curves on the models in the plot range. } 
\label{fig-rc-halos}
\end{figure} 

Decomposition of rotation curves by the $e^{-r^{1/4}}$ law surface mass profiles have been extensively applied to spheroidal components of late type galaxies (Noordermeer 2007, 2008). The $e^{-r^{1/4}}$ law was fully discussed in relation to its dynamical relation to the galactic structure based on a more general profile with $e^{-r^n}$ (Ciotti 1991; Trujillo 2002). However, there has been no attempt to apply it specifically to the Milky Way's rotation curve. 

In figures \ref{fig-rc-arm} and \ref{fig-2ring-hi}, we show the calculated rotation curve for the bulge model. The result shows a reasonable fitting of the inner rotation curve by a \dv-law bulge. Our result shows that the central steep rise and sharp peak at $R=300$ pc of the observed rotation curve is almost perfectly reproduced by the high density mass concentration in the central region of the bulge. 

The best-fit total mass of the bulge is as high as $1.8\times10^{10}\Msun$ and the scale radius is as compact as $R_b=0.5$ kpc with the accuracy of about 5\%. As figure \ref{fig-smd} indicates, the surface mass density in the central 0.5 kpc is dominated by the bulge component, and nuclear density reaches a value as high as $6.8 \times 10^6 \Msun{\rm pc}^{-2}$. The total projected mass included in the central 500 pc (scale radius) is $9.0 \times 10^9 \Msun$. The total spheroidal mass integrated in a sphere of $R_b$ is 0.39 times the total bulge mass, which is for the present case $7.0 \times 10^{9}\Msun$.

The surface mass distribution in figure \ref{fig-smd}  may be compared with the $K$ band surface brightness profile for the inner Galaxy at $| b |<10^\circ$ as presented by Kent et al. (1991, 1992). We now compare the integrated mass and luminosity for the central part of the bulge. The $K$ band luminosity profile for the central 0.94 kpc is expressed as  
\be
\nu_K=1.04 \times 10^6 (r/0.482 {\rm pc})^{-1.85} L_\odot {\rm pc}^{-3}. \label{Kent92}
\ee
This expression approximates the density profile for an isothermal sphere with a constant mass-to-luminosity ratio, for which the power-law index is $-2$. Since the functional form is different from that for the Vaucouleurs law as used here, we cannot compare the profiles directly. However, it may be worthy to compare the integrated luminosity within a radius with the corresponding mass. The integrated luminosity within a sphere of radius 0.5 kpc from equation (\ref{Kent92}) is calculated to be  $3.74 \times 10^9 L_\odot$. Thus, we obtain a mass-to-luminosity ratio for the bulge within a sphere of the scale radius, $R_b=0.5$ kpc, to be $M/L_{\rm bulge, 0.5 kpc}=7.1\Msun/L_\odot$.

As figure \ref{fig-smd} indicates, the volume density increases rapidly toward the Galactic Center, approaching an infinite value. The central mass within 1 pc is estimated to be as high as several $10^6 \Msun$. Such a high value is indeed observed near the nucleus (Kent 1992; Weiland et al. 1994). The high-density concentration at the center may be related to the central massive black hole of a mass $\sim 3\times 10^6 \Msun$ (Genzel et al. 1997, 2000; Ghez et al. 1998, 2000). 

\subsection{Exponential Disk}

The galactic disk is represented by an exponential disk (Freeman 1970). The surface mass density is expressed as
\be 
\Sigma_d (r)=\Sigma_{dc} {\rm exp}(-r/R_d)+\Delta, \label{eq_disk}
\ee 
where $\Sigma_{dc}$ is the central value, $R_d$ is the scale radius, and $\Delta$ is density perturbations such as due to arms, rings, and/or a bar as will be discussed below. The total mass of the exponential disk is given by $M_{\rm disk}= 2 \pi \Sigma_{dc} R_d^2$.

The rotation curve for a thin exponential disk is expressed by modified Bessel functions (Freeman 1970; Binney and Tremaine 1987). Here, we are interested in rotation curves affected by additional masses $\Delta$ due to arms, rings, bar, and/or interstellar gas. We, therefore, directly calculate the gravitational force $f(R)$ acting on a point at galacto-centric distance $x=R$ by integrating the $x$ directional component of force due to a mass element $\Sigma_d(r) dx dy$ in the Cartesian coordinates $(x,y)$:
\be 
f(R)=G \int_{-\infty}^\infty \int_{-\infty}^\infty 
{\Sigma_d (r) (R-x) \over s^3} dx dy ,
\ee
where $ s=\sqrt{(R-x)^2+y^2} $ is the distance between the mass element and the point.  The rotation velocity in order for the centrifugal force on a test particle at radius $R$ circularly rotating in the disk plane to balance with the gravitational force by the disk is thus calculated by $V_d(R) = \sqrt{f R}$.

\subsection{Dark Halo}

We assume a semi-isothermal spherical distribution for the dark halo (e.g.  Kent 1986). The density profile is written as
\be
\rho_h(r)=\rho_{hc} \left[ 1+\left( r \over R_h \right)^2\right]^{-1},
\label{eq_halo}
\ee
where $\rho_{hc}$ and $R_h$ are constants giving the central mass density and scale radius of the halo, respectively. This profile gives finite mass density at the center, but yields a flat rotation curve at large radius. The circular velocity is given by 
\be
V_h(r)=V_\infty \left[1-\left(R_h \over r \right) {\rm tan}^{-1}\left(r \over R_h \right) \right],
\ee
where $V_\infty$ is a constant giving the flat rotation velocity at infinity. The constants are related to each other as
\be
V_\infty=\sqrt{ 4 \pi G \rho_{hc} R_h^2},
\ee
or the central density is written as 
\be
\rho_0=0.740 \left(V_\infty \over {200 {\rm km ~s^{-1}}} \right) \left(R_h \over {1 {\rm kpc}} \right)^{-2}   \Msun {\rm pc}^{-3}.
\ee


Fitting to a dark halo model will be still crude not only because of the large scatter and errors of observed rotation velocities in the outer disk, but also for the weaker response of the rotation curve to the halo models. We here adopted the semi-isothermal halo as equation (\ref{eq_halo}) with a scale radius $R_h=5.5$ kpc and  flat rotation at infinity of $V_\infty=200$ \kms. However, we are able to obtain a reasonable fit for different sets of parameters, such as a higher $V_\infty$ and smaller $R_h$, keeping the nearly flat part between $R\sim 10$ and  20 kpc unchanged. Therefore, the fitted parameters to the halo component may not be unique. 

Reasonable fit is also obtained by other halo models such as the NFW (Navarro et al. 1995) and Burkert profiles (Burkert 1995).  Dehnen and Binney (1998) used similar profiles with a Gaussian truncation outside the outer galaxy. Any of these, including the isothermal model used here, yields smooth, nearly flat rotation curves at $R\sim 15-20$ kpc. In figure \ref{fig-rc-halos} we show the three rotation curves corresponding to the isothermal, Burkert and NFW models, where we see that the discrimination among the models may not be easy within the observed range up to 20 kpc. For any halo models with scale radius of about 5 to 10 kpc, the total mass enclosed within 20 kpc is about the same, $\sim 10^{11}\Msun$ (table \ref{tab_mass}).

The crudeness of fitting in the outer Galaxy, particularly at $R>10$ kpc, is mainly due to the large scatter of the observed data as well as  to large errors caused by uncertainties in the galacto-centric distances. We have recently one data point from VERA observations (Honma et al. 2007), which provides with galactocentric distance and velocity with errors as small as $\sim2$\% outside the solar circle. This data point is, however, significantly displaced from the mean velocities from the current observations.

We have recently combined the present rotation curve with the velocities of companion galaxies and dwarf galaxies in the Local Group to create a pseudo rotation curve for a the entire Local Group. We obtained a better fit by the NFW and/or Burkert model than isothermal (Sofue 2009).


\begin{figure*}
\begin{center}  
\includegraphics[width=16cm]{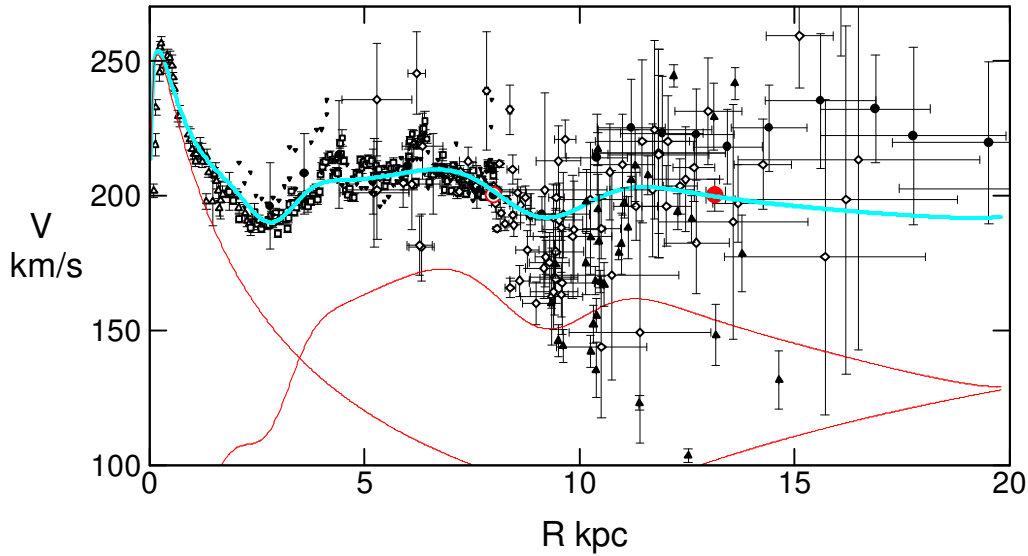} 
\end{center}
\caption{Model rotation curve compared with the observations. Thin lines represent the bulge, disk + rings, and dark halo components, and the thick line is the composite rotation curve. Data are the same as in figure \ref{fig-obs}.  } 
\label{fig-2ring-hi}
\end{figure*} 

\section{Perturbations}

In addition to the bulge, exponential disk, and dark halo, which have smooth and axisymmetric density distributions and construct the fundamental structure of the Galaxy, we then consider the effects of interstellar gas and local structure such as the bar, arms, and/or rings.

\subsection{Spiral arms}

We assume two-armed logarithmic spiral arms, whose density distribution is expressed as the following.
\be
\Delta(r,\theta)=\delta \Sigma_d(r) {\rm cos} \left[2 \left(\theta - {1 \over {\rm tan}~ p} {\rm log} {r \over a} -\alpha \right)\right], 
\label{eq-spiral}
\ee
where $\delta$ is the amplitude of the density waves, $p$ is the pitch angle of arms,  $a$ is a constant defining the position of an arm, and $\alpha$ adjusts arm's phase. Here, $r$ and $\theta$ are the cylindrical coordinates.

Figure \ref{fig-rc-arm}  shows a composite rotation curve for a model with spiral arms for $\delta=0.15$, $a=4$ kpc, $p=18^\circ$, and $\alpha=0^\circ$. The wavy characteristics of the observed curve are qualitatively reproduced, but the detailed fit is not sufficient. Particularly, the dip at $R\sim 9$ kpc cannot be explained by such spiral arms with usual density amplitude of $0.1 - 0.2$ times the background disk density. 
 
\subsection{Bar}

The effect of a bar on the gas kinematics is significant in the inner galaxy (Athanassoula 1992; Weiner and Sellwood 1992;  Fux 1997, 1999; Mulder and Liem 1986). The observations have indeed revealed a bar (Blitz et al. 1993; Blitz and Spergel  1991; Weiland et al. 1994; Freudenreich 1998). The rotation curve dip near 3 kpc and related detailed behavior may be discussed in the scheme of non-linear response of gas to the barred potential. In this paper, we examine only qualitatively if the 3 kpc dip manifests the existence of a bar. The central bar and noncircular flows within $\sim 150$ pc (e.g. Jenkins and Binney 1994) are beyond the resolution of the present rotation curve.

A bar is an extreme case of a spiral arm with a pitch angle $90^\circ$ in the above expression. Precisely speaking, circular rotation analysis cannot treat with non-axisymmetric potential. It is not a task here to evaluate the eight parameters of a bar: three axial lengths, position angle of major axis, mass, and density profiles along the axes, or at least six for a planar bar.  Orbit computation and galactic shock are also beyond the scope. 

Knowing such limitations, we estimate kinematical effect of a bar on the circular velocity. Since the half-mass radius of the bulge $R_{b}\sim 0.5$ kpc is much smaller than the observed bar length $\sim 2$ kpc (e.g. Freudenreich 1998), we treat the bar as a perturbation on the disk. We express the perturbation by equation (\ref{eq-spiral}) with $p\sim 90^\circ$, and the amplitude is replaced by 
\be
\delta=\delta_{bar} {\rm exp}[-(r/r_{bar})^2]
\ee
 with $r_{bar}$ being a cut-off radius. According to the COBE observations (Freudenreich 1998), we take $r_{bar}=1.7$ kpc and the tilt angle of $13^\circ$ from the Sun-Galactic center line. Since the bar width is small enough compared to $R_0$, we approximate that the lines of sight passes the bar side at a constant tilt angle of $15^\circ$, or $\alpha=75^\circ$. The amplitude was taken as $\delta_{bar}\sim$ 0.2 to 0.8. These mass profiles, together with the bulge and disk, approximately mimic the observed bar profiles.

Along an annulus at $r<\sim r_{bar}$ kpc, the circular velocity attains maximum on the bar side at  $r\sim 1$ kpc, where the density gradient is maximum. Faster velocities on the bar sides than on the major axis is indeed shown in the numerical computations in the literature as above. On the other hand, the density depression due to the sinusoidal perturbation causes slower gradient of potential at $r \sim 2$ kpc, leading to lower circular velocity at $\sim 3$ kpc. Our simple calculations showed enhancement of rotation velocity at $\sim 1$ kpc and a dip at 3 kpc, consistent with the observed rotation curve. However, the calculated dip amplitude was only a few \kms\ even for a large amplitude of $\delta_{bar}\sim 0.8$, which is too small to explain the observed dip. 

\subsection{Gaseous disk}

We examined the effect of gaseous disk observed in the HI and CO-line observations (Nakanishi and Sofue 2003, 2006). The surface mass fraction of gas is several \% in the inner disk, and it increases toward the outer disk, attaining several tens of percents at around 15 kpc. However, at these radii, the surface mass density of the dark halo much exceeds the disk as shown in figure \ref{fig-smd}. Since the circular velocity depends on the enclosed mass (bulge, disk, and halo), the contribution from the gaseous mass is negligible when computing the circular velocity at any radii.  

However, if we artificially multiply the gas density by 10 times, the wavy structure of rotation curve is mimicked, suggesting that the gas distribution is somehow related to more massive underlying structures. From these considerations, we may conclude that the kinematical effect of the gas disk is not so strong, and, therefore, the gas distribution is more passive, determined by the stellar and dark matter structures.

\subsection{Ring Waves}

In order to examine if the prominent dips in the observed rotation curve can be reproduced by local density enhancement and/or dips, we examine effect of wavy rings.  The rings are superposed on the exponential disk $\Sigma_d(r)$, so that the total density profile of the disk is expressed by the following equation:
\be
\Sigma_r=\Sigma_d(r) \left(1 +   \sum_{i=1}^2 f_i {\rm exp}(-t_i^2) {\rm sin}( {\pi \over 2} t_i) \right) , \label{eq_2ring}
\ee
where $f_i$ is the fractional amplitude of the $i$-th ring, $t_i=(r-r_i)/w_i$ with $r_i$ being the ring radius, and $w_i$ its width.

\begin{figure}
\begin{center}
\includegraphics[width=7cm]{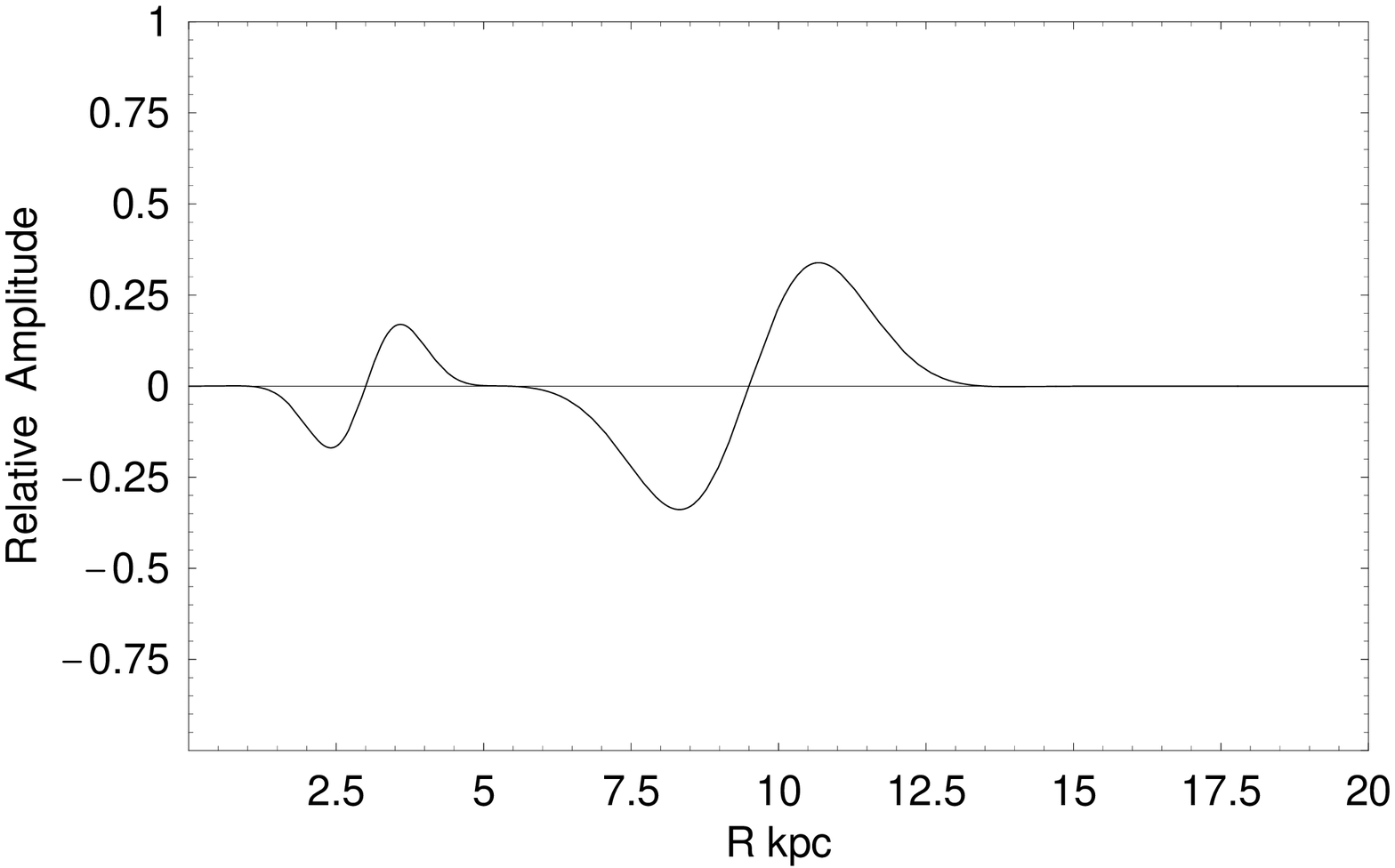} 
\includegraphics[width=7cm]{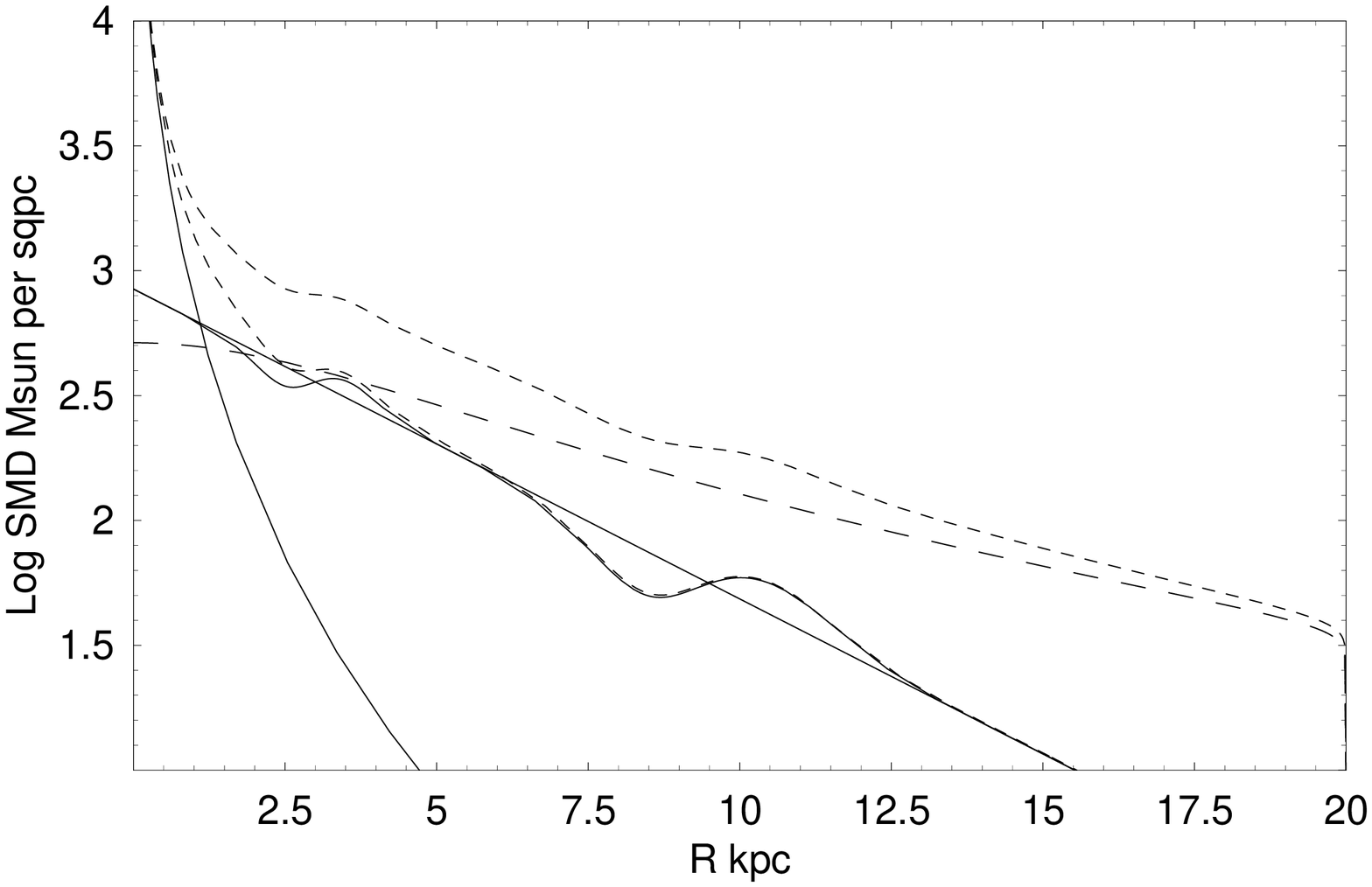}  
\end{center}
\caption{[Top panel]: Relative amplitudes of two wavy rings. [Lower panel]: Surface mass density of the disk superposed by the rings.  } 
\label{fig-2ring-smd}
\end{figure}

By the sinusoidal factor we represent a possible formation mechanism of the ring: We consider that the ring was produced from the same amount of mass swept up from inside the ring with the total mass being kept.  In order to obtain the best fit to the observations, we adopted the following values for the parameters as given in table \ref{tab_rings}. These parameters yield the maximum amplitude of $f=0.34$ for the outer ring. The profile of the surface mass density and relative amplitudes of the ring waves are shown in figure \ref{fig-2ring-smd}. The fitting result is discussed in the next section, and presented in figure \ref{fig-2ring-hi}. The bumpy features are fitted by  the ring wave model, where the surface mass density varies by about $\pm 0.17$ to 0.34 times the background exponential disk for the 3 and 9 kpc rings, respectively. Instead of the ring, the features may also be reproduced by introducing a spiral density wave of amplitude of about 20 \% of the exponential disk.

The 9 kpc dip requires a massive ring wave of node radius 9.5 kpc with the maximum at radius 11 kpc, minimum at 8.5 kpc, and the amplitude as high as $\sim 0.34$ times the underlying disk density. Since the rotation curve is based not only on the HI and molecular gases, but also on the observations of many stars (figures \ref{fig-obs}, \ref{fig-2ring-hi}), the 9 kpc dip is not considered to be due to some non-linear response of the gas on a weaker density wave. Hence, we conclude that there is indeed a ring-like density enhancement at $R=11$ kpc with its precursor dip at 8.5 kpc. We call such a massive ring wave the "great ring" at 11 kpc. Since the observed data are obtained for stars and gas within several kpc from the Sun, it is possible that this ring represents a density wave corresponding to the Perseus Arm (e.g. Nakanishi and Sofue 2003, 2006). The radial profile of the annulus-averaged HI density shows double peaks at 8 and 11 kpc. These HI peaks might be related to the great ring, but the HI kinematical properties are not particularly peculiar.

 
\subsection{Perturbations by Subhalos} 
It is possible that the halo itself is not spherical, but has substructures, which influence the dynamics of the galactic disk (e.g. Hayashi and Chiba 2006; Bekki and Chiba 2006). Such substructure would particularly affect the outer rotation curve. In fact the broad maximum seen in figure \ref{fig-obs} at $R\sim 15$ kpc could be due to such effect. However, inclusion of subhalo models is beyond the scope of this paper, and would be a subject for future numerical simulations. 

\begin{table}
\caption{Parameters for wavy rings expressed by equation (\ref{eq_2ring})} 
\begin{center}
\begin{tabular}{lllll}
\hline\hline
 \\ 
Ring $i$ & $r_i$ & $w_i$  & $f_i$ & $ \Sigma_r/\Sigma_d$ \\
&  (kpc) &  (kpc) &  &  (at peak) \\

\\ \hline \\

 1 & $r_1=3.0 $  & $w_1=1$ & $f_1=0.3$ &0.17 \\
2 & $r_2=9.5$ & $w_2=2$ & $f_2=0.6$ & 0.34 \\ 

\\
\hline \\ 
\end{tabular}
\end{center}
\label{tab_rings}
\end{table} 
 
\begin{table*}
\begin{center}
\caption{Parameters for Galactic mass components } 
\begin{tabular}{lllll}
\hline\hline   \\
Component & Parameter & Value 	& Uncertainty$^*$ \\
\\
\hline
\\
Bulge &Mass & 	$M_b=1.80\times10^{10}\Msun$ 	& $\sim 5$ \%\\
&Half-mass scale radius & $R_b=0.5$ kpc &  \\
&SMD at  $R_b$ &$\Sigma_{be}=3.2 \times 10^3 \Msun{\rm pc}^{-2}$ &  \\
&Center SMD & 	$\Sigma_{bc}=6.8\times10^6 \Msun{\rm pc}^{-2}$ &  \\
&Center volume density & $\rho_{bc}=\infty$ & ---	\\

\\ \hline  \\

Disk &Mass & $M_d=6.5\times10^{10} $ & $\sim 5$ \% \\
&Scale radius & $R_d=3.5$ kpc & \\
&Center SMD & $\Sigma_{dc}=8.44 \times 10^2 \Msun{\rm pc}^{-2}$ & \\
&Center volume density & $\rho_{dc}=8 \Msun{\rm pc}^{-3}$ & \\
\\ \hline \\

Rings &Mass & 	$M_r \sim 0$ &	\\
&Peak $\Sigma_r$ & 0.17 and 0.34 $\times \Sigma_d$ & $\sim20$  \% \\
&Radii of wave nodes & 	$R_r=3 $ and 9.5 kpc & $\sim3$ \\
&Widths & 	$ w_r=1$  and 2 kpc & $\sim10$ \\

Bar for 3 kpc dip 
&Amplitude $ \delta_{bar}$ & $>0.8 \times \Sigma_d$ & --- \\
&Assumed bar half length$^\dagger$ & 1.7 kpc & --- \\
&Assumed tilt angle$^\dagger$ & 	$13^\circ$ & --- \\

\\ \hline \\
 
Bulge, disk, rings &  Total mass & $M_{bdr}=8.3\times10^{10}\Msun$ & $\sim5$ \% \\

\\ \hline \\
 
Dark halo 
&Mass in $r=10$kpc sphere & $M_h(10{\rm kpc})=4.2\times10^{10}\Msun$ & $\sim10$ \%\\
(Spherical, isothermal)& Mass in $r=20$ kpc sphere$^\ddagger$  & $M_h(20{\rm kpc})=1.24 \times 10^{11} \Msun$ & \\
&Core radius	& $R_h=5.5$ kpc &  \\
&Central SMD in $|z|< 10$ kpc& $\Sigma_{hc}=352 \Msun{\rm pc}^{-2}$ & \\
  &Central volume density & $\rho_{hc}=0.03 \Msun{\rm pc}^{-3}$ &  \\
  &Circular velocity at infinity  & $V_\infty=200 {\rm km ~ s}^{-1}$ &(fixed)  \\
\\ \hline \\
Total Galactic mass & Mass in $r=20$ kpc sphere  & $ M_{\rm total}(20 {\rm kpc})=2.04 \times 10^{11} \Msun $ & $\sim10$ \% \\
\\ \hline
\end{tabular} 
\end{center}
$*$ Eye estimates after trial-and-error fitting of calculated rotation curve to the observations. \\
$\dagger$ Freudenreich (1998)\\
$\ddagger$ Mass within 20  kpc is weakly dependent on the dark halo models, e.g., about the same for the NFW and Burkert models.
\label{tab_mass}
\end{table*}

\begin{table*}
\caption{Local values in the solar vicinity at $R=R_0=8.0$ kpc.} 
\begin{center}
\begin{tabular}{lllll}
\hline\hline  \\  
& Components & Local values   \\
\\
\hline 
\\
Surface Mass Density & Bulge (\dv)  &  1.48 $\Msun {\rm pc}^{-2}$ \\
& Disk (exponential) &  87.5 $\Msun {\rm pc}^{-2}$ \\
& Dark halo (isothermal, $|z|<10$ kpc)&193 $\Msun {\rm pc}^{-2}$ \\ 
\\
\hline
\\
Volume Mass Density & Bulge  & $1.3 \times 10^{-4}$ $\Msun {\rm pc}^{-3}$\\
& Disk$^\dagger$ for $z_0=144$ pc & 0.30 $\Msun {\rm pc}^{-3}$ \\ 
&  --- for $z_0=247$ pc & 0.18 $\Msun {\rm pc}^{-3}$ \\ 
& Dark halo    & $9.6 \times 10^{-3}$  $\Msun {\rm pc}^{-3}$\\
& Sum & $0.19 \sim 0.31 \Msun {\rm pc}^{-3}$
\\
\hline
\\
Total Mass within Solar Sphere$~\ddagger$ & Bulge in sphere of $r=R_0=8$ kpc &  $1.75\times10^{10}\Msun$ \\
& Disk in $R=R_0$  & $  4.33\times10^{10}\Msun$ \\
& Dark halo in sphere $r=R_0$ &  $8.31\times10^9\Msun$ \\
& Total mass$^*$ in sphere of $r=R_0$ &  $6.91\times10^{10}\Msun$ \\
\\
\hline
\end{tabular} \\
\end{center}
$\dagger$ For scale heights $z_0=247$ pc (Kent et al. 1991) and 144 pc (Kong and Zhu 2007).\\
$\ddagger$ The "Solar sphere" is a sphere of radius $R_0=8$ kpc centered on the Galactic Center.\\
$*$ Slightly smaller than the Keplerian mass $7.43\times10^{10}\Msun$ for $V_0=200$ \kms because of the disk effect.

\label{tab_local}
\end{table*}

\section{Discussion}

We have obtained an updated rotation curve by integrating the current data in the decades, and plotted them in the same scale for a nominal galactic constants $R_0=8$ kpc and $V_0=200$ \kms. We have decomposed the obtained unified rotation curve into a bulge, disk and dark halo components. The functional form of the bulge was so adopted that the surface mass density is represented by the \dv\ law, which was tried for the first time for our Galaxy. The disk was approximated by an exponential disk, and the halo by an isothermal sphere. The observed characteristics are well fitted by superposition of these components. The central steep rise and the high rotation peak at $R= 300$ pc is quite well reproduced by the \dv\ bulge of half-mass scale radius $R_b=0.5$ kpc. The broad maximum at around $R\sim 6$ kpc was fitted by the exponential disk, and the flat outer part by a usual dark halo. 

Table 3 lists the fitting parameters for individual mass components. Since the used data in figure \ref{fig-obs} were compiled from different observations, their errors are not uniform, and it was not straightforward to estimate the statistical errors by calculation. It was particularly difficult in the outer Galaxy, where we gave the highest priority to the observation by VERA (Honma et al. 2007): The other data, largely scattered around the mean values, were referenced to judge if the fitting result was reasonable. So, we give only eye-estimated values as evaluated after trial and error of fitting to the observed points. Nevertheless, one may be interested in an averaged rotation curve from the current data, and we show it by the thick dashed line in figure \ref{fig-rc-arm}. It indicates a simply averaged observed rotation curve taken from Sofue et al. (1999), where the outer curve is based only on the HI data of Honma and Sofue (1997a), but the data from OB stars and C stars are not included for their particularly large scatter mainly due to the uncertainties of distance determinations.

The local values of the surface mass and volume densities in the solar vicinity  calculated for these parameters are also shown in table \ref{tab_local}. The  volume density of the disk has been calculated by $\rho_d=\Sigma_d /(2 z_0)$ with $z_0$ being the scale height at $R=R_0$, when we approximate the disk scale profile by $\rho_d(R_0, z)=\rho_{d0}(R_0) {\rm sech} z/z_0$. For the local thin galactic disk, we adopted a recent value  $z_0=144\pm10$ pc for late type stars based on the Hipparcos star catalogue (Kong and Zhu 2008). The local volume density of the bulge is four orders of magnitudes smaller than the disk component, and the halo density is two orders of magnitudes smaller. However, the surface mass densities as projected on the Galactic plane are not negligible. The bulge contributes to 1.6\% of the disk value, or the stars in the direction of the galactic pole would include about 2\% bulge stars, given the  \dv\ density profile. The dark halo mass integrated within heights of $-10<z<10$ kpc exceeds the disk value by a factor of 2.1.

Shorter-scale variations are superposed on the rotation curve. The deep minima at $3$ and 9 kpc are the most prominent perturbations. These features are not reproduced by the basic mass components. The 3 kpc dip is consistent with the bar observed with COBE. It was also possible to fit it by adding a wavy ring of radius 4 kpc (node at 3 kpc), which may be related to the dense molecular ring of radius 4 kpc. The most striking, peculiar feature firmly confirmed in the unified present rotation curve is the 9 kpc dip. We have attributed this dip to a massive ring at 11 kpc, which we called the great ring. The required density perturbation is as large as  $\pm$0.34 times the underlying disk density. This is much deeper than that expected for spiral density waves of $\sim 0.1-0.2$. However, the HI gas in the Perseus and nearby arms seems to be not as strongly disturbed as expected from the great ring.
 

Finally, we comment on the original question whether it is indeed not possible to explain the observed rotation curve, including the 9 kpc dip, only with a bulge, disk, and a halo. This is part of the disk halo conspiracy, in which the disk and halo conspire to create a flat rotation curve, but we might see here at the dip an indication of a transition between halo and disk. However, such transition seems difficult to be recognized on the theoretical rotation curves, even if it existed, because of the long-range force of the gravity as well as the smooth distributions of the disk and halo masses. One may be convinced with this from the model calculations in figure \ref{fig-rc-arm}. This is in fact the reason why we had to attribute the 9-kpc dip to local structures in the disk.

\vskip 5mm

Acknowledgements: The authors are indebted to the anonymous referee for the valuable comments and suggestion to add: discussion of reality of the 9 kpc dip related to the possibility of pressure support of C-star disk by high velocity dispersion; the effect of subhalos on the outer rotation curve; and the disk-halo conspiracy in creating a flat rotation curve.


{}   

\end{document}